\documentclass[aps,preprint]{revtex4}%
\usepackage{amsfonts}
\usepackage{amsmath}
\usepackage{amssymb}
\usepackage{graphicx}%
\begin{document}
\title[Traffic clearance distributions]{Modelling highway-traffic headway distributions using superstatistics}
\author{A. Y. Abul-Magd}
\affiliation{Faculty of Engineering, Sinai University, El-Arish, Egypt}

\begin{abstract}
We study traffic clearance distributions (i.e., the instantaneous gap between
successive vehicles) and time headway distributions by applying Beck and
Cohen's superstatistics. We model the transition from free phase to congested
phase with the increase of vehicle\ density as a transition from the Poisson
statistics to that of the random matrix theory. We derive an analytic
expression for the spacing distributions that interpolates from the Poisson
distribution and Wigner's surmise and apply it to the distributions of the
nett distance and time gaps among the succeeding cars at different densities
of traffic flow. The obtained distribution fits the experimental results for
single-vehicle data of the Dutch freeway A9 and the German freeway A5.

\end{abstract}
\volumeyear{year}
\volumenumber{number}
\issuenumber{number}
\eid{identifier}
\date[Date text]{ \today}
\startpage{1}
\endpage{2}
\maketitle

In empirical highway traffic observations \cite{chowdhury}, it has been found
that traffic can be either free or congested. In free flow conditions, drivers
can choose their own speed. The distances between cars are uncorrelated, and
thus follow a Poisson distribution. Traffic congestion is a road condition
characterized by slower speeds, longer trip times, and increased queueing. It
occurs when traffic demand is greater than the capacity of a road (or of the
intersections along the road). Kerner \cite{kerner} classified the congestion
regime into two distinct phases: synchronized flow and wide moving jams. In
synchronized flow, the speeds of the vehicles are low and vary quite a lot
between vehicles, but the traffic flow remains close to free flow. In wide
moving jams, vehicle speeds are more equal and lower, and time delays can be
quite large. Extreme traffic congestion, where vehicles are fully stopped for
periods of time, is colloquially known as a traffic jam. Besides Kerner's
'three-phase theory', congested traffic has been described in terms of five
congestion phases of different spatiotemporal properties, and their
combinations \cite{helbing1}. A velocity dependent randomization variant of
traffic cellular automata leads to the emergence of four distinct phases: free
flowing traffic, dilutely congested traffic, densely advancing traffic, and
heavily congested traffic \cite{maer}.

The question whether the transition from one regime to another is a smooth
crossover or is a result of a genuine phase transition is still not settled in
most traffic models. Measurements of traffic breakdown on various highways by
Kerner and Rehborn \cite{kerner1} indicate a first-order transition between
free flow and synchronized traffic. Some traffic cellular automata models are
suggested to exhibit phase transitions (see, e.g. \cite{schad} and references
therein). However, the existence of such a transition can only be explicitly
demonstrated in some limiting cases where certain dynamical processes are
deterministic. Asymmetric chipping models, where a single particle hops to a
nearest neighbor site at a constant rate, suggest that the jamming phase
transition does not take place. Rather, the system exhibits a smooth crossover
between free-flow and jammed states, as the car density is increased
\cite{levine}.

Krb\'{a}lek and \v{S}eba studied the statistics of bus arrivals close to the
Cuernavaca city center \cite{seba} where the distances between busses is
optimized. They found that the Gaussian unitary ensemble (GUE) of
random-matrix theory (RMT) \cite{mehta}%
\begin{equation}
P_{\text{GUE}}(s)=\frac{32}{\pi^{2}}\frac{s^{2}}{D^{3}}e^{-4s^{2}/\pi D^{2}},
\end{equation}
where $D$ is the mean spacing, successfully models the bus spacing
distribution, and also the bus number variance measuring the fluctuations of
the total number of buses arriving at a fixed location during a given time
interval. Krb\'{a}lek and \v{S}eba justified the GUE properties of the bus
arrival statistics by regarding the buses as one dimensional interacting gas.
Traffic is treated as a gas of interacting particles (vehicles) described by a
distribution function with time evolution given by a Boltzmann equation. The
steady state solution is given by the Boltzmann factor. The probability
density function for the positions of the charges is given by $\exp(-\beta V)$
times a constant, where $V$  and $\beta~$are the potential energy and\ the
inverse temperature of the gas, respectively. Dyson has shown several years
ago that the exact level statistics of random matrix ensembles is obtained for
Coulomb interaction between the gas particles, when the car positions are
identified with the eigenvalues of the ensemble matrices \cite{mehta}. GUE
corresponds to a value of $\beta=2.$ Abul-Magd \cite{abul1} found that the
spacing distribution of GUE agrees with data measuring the gaps between parked
cars on four streets in central London \cite{rawal}, where it is difficult to
find a parking place. In the light of these empirical findings, we expect GUE
to reasonably describe traffic in the phase of wide (moving) jams, which are
localized structures propagating upstream \cite{kerner} with a mean spacing
$D$ between vehicles close to the minimum (safe) gap.

We shall therefore regard formula (1) as an empirical result for jammed
traffic. On the other hand, the distances between cars if free-flow traffic
are uncorrelated, and the spacing distribution follow the Poisson distribution%
\begin{equation}
P_{\text{Poisson}}(s)=\frac{1}{D}e^{-s/D}.
\end{equation}
Other forms of congested traffic that constitute synchronized flow
\cite{kerner}, whose main characteristic is the apparent absence of a
functional flow-density form, may be considered as a transition between
(almost) regular dynamics and chaos. In this paper, we describe them using the
concept of superstatistics (statistics of a statistics).

Superstatistics has been applied to model systems with partially chaotic
classical dynamics within the framework of RMT in Ref. \cite{supst,sust}. The
formalism of superstatistics has been proposed by Beck and Cohen \cite{BC} as
a possible generalization of statistical mechanics. The superstatistics
concept is very general and has been applied to a variety of complex systems
(see \cite{beck} and references therein). Essential for the superstatistical
approach is the existence of an intensive parameter, which fluctuates on a
large spatiotemporal scale compared to the fluctuations of the constituents of
the system. In congested traffic, there is a relatively fast dynamics given by
the velocity of the vehicle and a slow one given by the traffic density, which
is spatiotemporally inhomogeneous. The two effects produce a superposition of
two statistics, i.e. superstatistics. In fact, a traffic headway is
inhomogeneous in space and in time. Effectively, it may consist of many
spatial cells, where there are different values of the traffic density. Here
the flow will always increase (decrease) with increasing density. In
time-series of flow-density measurements, the flow might increase or decrease,
in sharp contrast to what is observed for the free-flow and jammed phase. The
measured time series may consist of many time slices. This \textquotedblleft
irregularity\textquotedblright\ of the time-series has been quantified by
using the cross-correlation function \cite{neubert} between density and flow.
It is very close to 1 in free flow and the jams but almost 0 for synchronized
flow. The order parameter introduced in\ \cite{lubashevsky} to reflect \ the
degree of the internal interaction of vehicles takes a large value for
synchronized flow, whereas free flow and the jam match its small values as
weak mutual interactions.

Within the superstatistics framework, the car ensemble compromises various
groups of cars jammed with different "local" mean spacing. We express\ any
statistic $P$ of a (sufficiently chaotic) congested traffic as an average of
the corresponding statistic $P^{(G)}(D)$ for a Gaussian random ensemble over
the local mean level spacing $D$. The superstatistical generalization is given
by
\begin{equation}
P=\int_{0}^{\infty}f(D)P^{(G)}(D)dD.
\end{equation}

Our goal in the present paper is to show that nearest-neighbor-spacing
distribution (NNSD) obtained by substituting $P_{\text{GUE}}(s)$ for
$P^{(G)}(D)$ in Eq. (3), provides a plausible explanation for the observed
car-spacing distribution at arbitrary traffic densities. The form of the
probability distribution $P$ as a weighted sum over equilibrium distributions
$P^{(G)}$ up to now is based on purely phenomenological arguments. No
fundamental derivation of Eq. (3) from the basic principles has been given. In
this case the main problem is the relationship between the weights $f(D)$ of
this expansion and specific random processes governing the system dynamics,
which is currently the main direction of researches carried out in the field
of superstatistics (see, e.g., Ref. \cite{lubashevsky1}). Following Sattin
\cite{sattin}, we evaluate $f(D)$ by using the principle of maximum entropy.
Lacking a detailed information about the mechanism causing the deviation from
the prediction of RMT, the most probable realization of $f(D)$ will be the one
that extremizes the Shannon entropy $S=-\int_{0}^{\infty}f(D)\ln f(D)dD$ with
the following constraints: (i) Basic for most statistical applications is the
distinction between average quantities and their fluctuations. The fluctuation
properties are defined in RMT for unfolded spectra, which have a unit mean
level spacing. We thus require $\int_{0}^{\infty}f(D)DdD=1.$\textbf{(}ii)
Vehicle headway distributions are measure of narrow intervals of traffic
densities. We require that mean level-density of the superposed GUE's
$\left\langle D^{-1}\right\rangle =\int_{0}^{\infty}f(D)D^{-1}dD$ is fixed.
With these constraints, the maximization of $S$ yields
\begin{equation}
f(D)=C(\alpha,D_{0})\exp\left[  -\alpha\left(  \frac{D}{D_{0}}+\frac{D_{0}}%
{D}\right)  \right]  ,
\end{equation}
where $\alpha$ and $D_{0}$ are parameters, which can be expressed in terms of
the Lagrange multipliers of the constrained extremization, and $C(\alpha
,D_{0})=1/2D_{0}K_{1}\left(  2\alpha\right)  $ where $K_{m}(x)$ is a modified
Bessel function of the second kind. The parameter $D_{0}$ is given by the
condition of $\left\langle D\right\rangle =1$ as%
\begin{equation}
D_{0}=\frac{K_{1}\left(  2\alpha\right)  }{K_{2}\left(  2\alpha\right)  }.
\end{equation}
The parameter $\alpha$ defines the dispersion of the local mean spacing, whose
variance is given by%
\begin{align}
\sigma^{2}  &  =\frac{K_{1}\left(  2\alpha\right)  K_{3}\left(  2\alpha
\right)  }{\left[  K_{2}\left(  2\alpha\right)  \right]  ^{2}}-1\nonumber\\
&  \approx\left\{
\begin{array}
[c]{c}%
1+\{0.536274+4\ln(2\alpha)\}\alpha^{2}+O(\alpha^{3})\text{, for small }%
\alpha,\\
\frac{1}{2\alpha}+O\left(  \frac{1}{\alpha^{3}}\right)  \text{, for large
}\alpha.
\end{array}
\right.  .
\end{align}
The distribution (4) tends to $\delta(D-1)$ at large $\alpha$, which
correspond to traffic jams by assumption.

We note that the constraint (ii) imposed here is different from the
corresponding constraint used in \cite{sust}, which requires the existence of
$\left\langle D^{-2}\right\rangle .$ Namely the present choice has recently
been found \cite{abul2} to produce a distribution of local density of states
$\nu=1/D$ which is very similar to the one obtained by Altshuler and Prigodin
\cite{altsher} for strictly one-dimensional disordered chains, which has been
successfully applied in the numerical simulation of the closed wire (see,
e.g., \cite{schomerus}). In addition, the present constraint (ii) is more
suitable for analysis of traffic data, such as the data to be considered in
this paper, where vehicle spacings are taken from separate intervals of
traffic density.

We now apply superstatistics to calculate the NNSD for a system undergoing a
transition out of chaos described by GUE statistics. For this purpose, we
substitute Eqs. (5,6) into (4) to obtain%
\begin{equation}
P(\alpha,s)=\frac{16}{\pi^{2}D_{0}K_{1}\left(  2\alpha\right)  }s^{2}\int
_{0}^{\infty}\exp\left[  -\alpha\left(  \frac{D}{D_{0}}+\frac{D_{0}}%
{D}\right)  -\frac{4s^{2}}{\pi D^{2}}\right]  \frac{dD}{D^{3}}%
\end{equation}
where $D_{0}$ is given by Eq. (5).

We now try to model the transition of traffic from the free flow regime to
that of a moving jam as a dynamical transformation from the Poisson statistics
to that of a GUE. Accordingly, the traffic headway NNSD at intermediate
traffic densities has an intermediate behavior between the Poisson and GUE
distributions (3) and (5). A quantitative interpretation of this model is
provided by the superstatistical generalization of RMT. In the following, we
show that the superstatistical spacing distribution in Eq. (7) is suitable for
describing clearance distribution at arbitrary traffic density.

The distributions of space gap between vehicles (i.e. clearances in the
traffic terminology) are recorded by double induction-loop detectors
continuously during approximately 140 days on the Dutch two-lane freeway A9
\cite{wagner}. The macroscopic traffic density was calculated for samples of
$N=50$ subsequent cars passing a detector. The region of the measured
densities $\rho$ $\in$ [0, 85 vehicle/km/lane] is divided into 85 equidistant
subintervals. The measured data in each density subinterval include, for each
lane, the passage time of \ each vehicle, its velocity and its length. From
this, the individual bumper-to-bumper distance $s_{i}$ among the succeeding
cars ($i$th and $(i-1)$th) are determined. The bumper-to-bumper distance
$s_{i}$ among the succeeding cars ($i$th and $(i-1)$th) is calculated (after
eliminating car-truck, truck-car, and truck-truck gaps). The mean distance
among the cars is re-scaled to 1 in all density regions. The car-spacing
distributions of four density regions, which have been reported in Ref.
\cite{krbalek} are shown in Fig. 1 by histograms. The curves are the best fit
to the superstatistical distribution (7). The figure demonstrates the high
quality of agreement between the proposed model and the experiment. The
best-fit values of the superstatistical parameter $\alpha$ are 0.16, 0.48, 2.1
and 9.3, for density intervals centered around $\rho=0.5,4.5,25.5~$and 81.5
vehicle/km/lane, respectively. Interestingly, if we disregard the first
interval where the spacing distribution is nearly Poissonian, we find that the
best-fit values of $\alpha$ increases linearly with $\rho$ such that%
\begin{equation}
\alpha=c\rho,~c=0.10\pm0.1.
\end{equation}

In Ref. \cite{krbalek}, the empirical clearance distributions are successfully
compared with a one-dimensional thermodynamical particle gas model. The author
considers a system of identical particles on the circumference of a circle.
The particles interact with a repulsive potential inversely proportional to
their mutual distance. The agreement between experimental and calculated
distributions is obtained by varying one free parameter (inverse temperature
$\beta$) that represents the traffic density. In spite of the success of this
model, it is not obvious to us that (equilibrium) thermostatics can describe
all of the different traffic phases in a similar way.

The second quantity we look at is the time headway distribution, which is the
time elapsing between two vehicles passing the detector. In principle, time
headways are associated with space headways once the latter are measured in
narrow vehicle-density intervals. The required single-vehicle data include,
for each lane, the passage time$~t_{i}^{0}$ of vehicle $i$, its velocity
$v_{i}$, and its length $l_{i}$. From this, we determine the individual netto
time gaps as $t_{i}=t_{i}^{0}-t_{i-1}^{0}-l_{i}/v_{i}$. Kerner et al.
\cite{kerner2} measured single vehicle data sets on the three-lane freeway
section of the German freeway A5-South. They reported the time headway
distributions for the free-flow, synchronized flow and moving jam phases. We
calculated the mean values $\overline{t}$ of the netto time gaps in these
distributions to be 1.52, 1.82 and 2.29 s, respectively. The distributions
$P(\tau)$, where $\tau=t/\overline{t},$ are compared in Fig. 2 with the
superstatistical distribution in Eq. (7). The agreement between the empirical
and superstatistical distributions is very good. The best-fit values of the
parameters $\alpha$ take large values $(=$ 5.9, 15 and 13 respectively)
compared to those obtained in fitting \ the headway data for the same traffic
phase. The corresponding values of the variance $\sigma^{2}$ of the parameter
distribution $f(D)$\ are quite small, being 0.084, 0.033 and 0.038,
respectively. Due to the slow decrease of the variance $\sigma^{2}$ at large
values of $\alpha$ (see last line of Eq. (6)), the distributions $P(\tau)$
look very similar for all the three phases. A possible reason for the large
values if $\alpha$ is the additional fluctuation of the mean time headway
introduced by the mean-velocity fluctuation to that of the local mean space
gap $D$. Surprisingly, the best fit value of the superstatistical parameter
$\alpha$ is almost the same for both synchronized flow and the jam
($\alpha=15,13$ respectively), both have almost a GUE distribution.

To summarize, the space-gap distribution between vehicles in traffic jams
shows strong "level repulsion" and is well reproduced by the Wigner surmise
for GUE. The clear distances between vehicles in a free-flowing traffic, on
the other hand, are uncorrelated and follow Poisson statistics. In this paper
we use the concept of superstatistics to model congested traffic as a
superposition of moving jams represented by GUE's with different mean level
spacings. We derive an expression for NNSD that describes the transition
between the Poisson-like statistics to that of a GUE by tuning a single
parameter, namely the superstatistical parameter $\alpha$. This parameter
measures the variance of the fluctuating intensive variable (the mean distance
between vehicles). We then apply the derived distribution to model transition
of traffic from a stationary free-flow phase to a continuously growing
congested non-stationary phase. Small values of $\alpha$ correspond to the
free- flowing traffic, while the jammed and congested traffic phases are
described by spacing distributions with large values of $\alpha$. We found
that the superstatistical NNSD provided a satisfactory description for the
distance headway distributions at different densities by varying a single
parameter. Thus, that the statistical features of traffic clearance exhibit a
smooth crossover between a free-flow and a jammed state as the car density is
increased. The best-fit values of superstatistical parameter increases almost
linearly with the traffic density. This may suggest that single-vehicle data
do not "feel" a first-order phase transition in traffic flow. However, it is
not possible to draw a definite conclusion by considering so small number of
cases. We note that the values of the inverse temperature of the
one-dimensional gas model that fitted these and many other similar data
\cite{krbalek} also do not show any step-wise variations as the car density
increases. The superstatistical distribution was also successfully applied to
time-headway distributions measured by Kerner and collaborators for the
free-flow, synchronized flow and moving jam traffic phases. While the mean
time gaps between vehicles were essentially different in the three traffic
phases, the fluctuation properties (measured by NNSD of "unfolded" levels)
were almost the same for the two congested phases but rather different from
the case of free flow.

The presented results support the possibility for applying the
superstatistical RMT to the traffic systems. Obviously, the proposed model is
no substitute for the elaborate investigations of the traffic problems.
Nevertheless, the powerful methods of RMT may be useful in understanding some
of its aspects even during the transition between the free-flow and congested phases.

\pagebreak

\bigskip{\Large Figure Caption}

Figure 1. Probability density $P$($s$) for scaled spacing $s$ between
successive cars in traffic flow, taken from Ref. \cite{krbalek}. Histograms
represent the clearance distributions computed for traffic data from four
density region, $\rho\in(0,1),(4,5),(25,26),(81,82)$ (in vehicle/km/lane). The
curves represent the predictions of superstatistical model (8) where the
best-fit values of the superstatistical parameter $\alpha$ are 0.16, 0.48, 2.1
and 9.3, respectively.

Figure 2. Probability density $P$($\tau$) for scaled netto time gaps $\tau$
between successive cars in traffic flow, taken from Ref. \cite{kerner2}.
Histograms represent the time-headway distributions computed for traffic data
for the free-flow, synchronized flow and moving jam traffic phases. The curves
represent the predictions of superstatistical model (8) where the best-fit
values of the superstatistical parameter $\alpha$ are 5.9, 15 and 13 respectively.

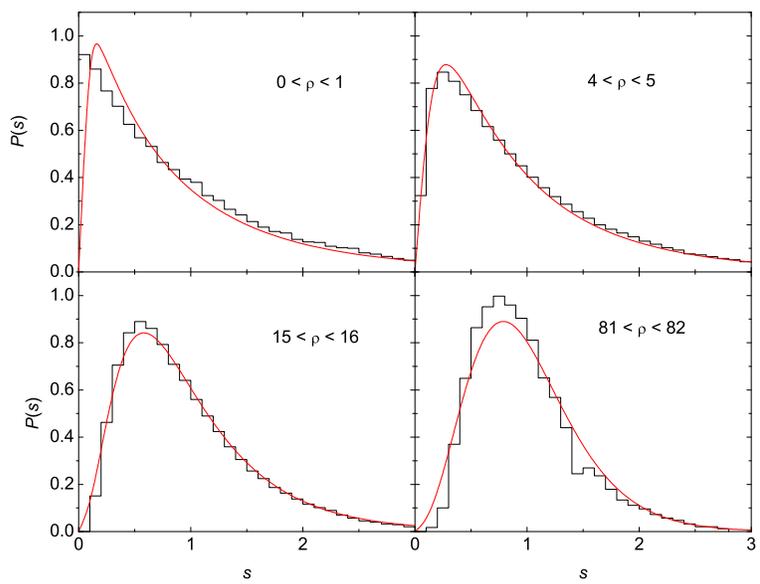
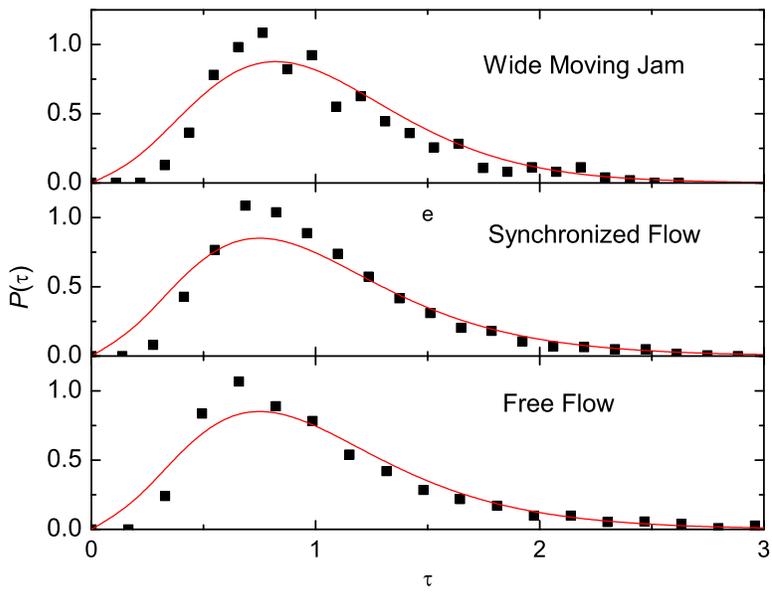
\end{document}